\documentclass[10pt,twoside,twocolumn]{article}
\usepackage[latin1]{inputenc}
\usepackage[T1]{fontenc}
\usepackage{graphicx}
\usepackage{hyperref}
\hypersetup{colorlinks=true, linkcolor=blue, filecolor=blue, pagecolor=blue, urlcolor=blue}

\usepackage[vmargin=20mm,hmargin=24mm]{geometry}\usepackage{cmbright}
\def\auteur{Bernard \textsc{Jacquemin}}
\def\hbpauteur{Bernard \textsc{Jacquemin}}
\def\adresselabo{Université de Haute Alace / CREM EA3476\\10 rue des Frères Lumière F68\,093 Mulhouse cedex (France)}

\def\courriel{\href{Bernard.Jacquemin@uha.fr}{Bernard.Jacquemin\,@\,uha.fr}}

\def\titre{A derivational rephrasing experiment for question answering}

\def\titrecourt{\titre}

\def\piedpage{LREC 2010 \textit{(Sous presse)}}

\usepackage{fancyhdr}
\title{\titre}
\author{\auteur\\\adresselabo\\\courriel}
\date{}

\usepackage{pstricks,pst-node,pst-text,pst-3d}

\begin{document}

\maketitle

\begin{abstract}
 In Knowledge Management, variations in information expressions have proven a real challenge. In particular, classical semantic relations (e.g. synonymy) do not connect words with different parts-of-speech. The method proposed tries to address this issue. It consists in building a derivational resource from a morphological derivation tool together with derivational guidelines from a dictionary in order to store only correct derivatives. This resource, combined with a syntactic parser, a semantic disambiguator and some derivational patterns, helps to reformulate an original sentence while keeping the initial meaning in a convincing manner This approach has been evaluated in three different ways: the precision of the derivatives produced from a lemma; its ability to provide well-formed reformulations from an original sentence, preserving the initial meaning; its impact on the results coping with a real issue, \textit{ie} a question answering task. The evaluation of this approach through a question answering system shows the pros and cons of this system, while foreshadowing some interesting future developments.
\end{abstract}

\section{Introduction}

With the exponential increase of available textual documents, it has become impossible for anyone to read all of them, or manage all the information they contain. Automatic methods are thus necessary to deal with these masses of text and to provide quick and easy access to a piece of information lost in data. Among Knowledge Management disciplines that try to solve this issue, the question answering task [which consists in supplying the text phrase that contains the answer to a question] is particularly fussy: on the one hand the answer supplied has to be as concise and precise as in Information Extraction, and on the other, the system must adapt to the varying queries and address changing information types in order to find an answer, as does Information Retrieval. The major obstacle with which the question answering task is confronted consists in identifying text meaning: a difficult job for a computer. The same piece of information is indeed phrased in different ways in a question and in the questioned text base. These differences prevent the system from matching data and consequently from extracting the right answer \cite{GrauMagnini05,StrzalkowskiHarabagiu06}.

Several approaches have been proposed to tackle this problem. Some of them attempted, and sometimes succeeded, in building semantic representations for query and textual utterances that were next matched \cite{GroisWilkins05,HarabagiuHickl06}. But the query expansion method, although simpler to carry out, is a very common choice in the discipline, because it covers a large amount of different phrasing of the same piece of information  \cite{GrauAl06,DangAl06}. The process consists generally in constituting, for each significant word in the query, a disjunctive list of terms with the same meaning as the original word. In order to find equivalent terms, classical semantic relations make it possible to draw up lists of synonyms, hyperonyms, etc. But these semantic relations do not give the opportunity to extend the rewording beyond the limits of the part-of-speech of the original word, and even more so to explore new syntactic schemata.

In order to free themselves from this part-of-speech constraint, many researchers have followed the morphological derivation trail, considering that members of the same derivative family have roughly the same meaning \cite{Church95,Jacquemin96,Hull96}. Nevertheless, the results reached by morphological derivation in a query expansion task are often inconclusive. Far from improving the quality and precision of answers, derivation systems tend to provoke numerous incorrect answers. At present, the current derivation systems are not able to generate the whole derivation family of a given word, without generating simultaneously several words incorrectly associated with the authentic derivatives, but morphologically, and above all semantically, distinct from them. On the contrary, some parameters and constraints can be defined for these generation tools in such a way that priority is given to precision, minimizing noise and eliminating scoria from \textit{candidates-derivatives}. But if this method significantly reduces the error rate, it also entails a dramatic reduction in recall that dispels the query expansion interest for textual information management \cite{GaussierAl00,BilottiAl04}.

Despite this assessment, a method that uses both a morphological derivation tool and a general French dictionary is proposed in order to build a rich and accurate derivational resource by filtering candidates-derivatives with derivational instructions. To take advantage of the just-generated derivatives in query expansion, parse the utterance is parsed and a Word Sense Derivation system applied \cite{JacqueminAl02}. The tool used to generate candidates-derivatives and the dictionary with the filtering instructions are presented herein, as are the means of constructing the derivational resource and a short evaluation of its quality; thereafter, the approach to the formulation of derivational rephrasings as close as possible to the original utterance meaning are outlined. Finally, the derivative rephrasing approach in the absolute and in terms of its impact on our question answering system's performance are evaluated.

\section{From generation to filtering of derivatives}

The method proposed consists first in generating as many words as possible that are likely to belong to the same derivational family as a given term in such a way as to get as many actual derivatives as possible among the candidates, whilst not taking into account the number of incorrect creations. Then all the inaccurate candidates-derivatives are excised from the list by filtering all the propositions that do not match the derivational instructions from the dictionary.

For many years, research has been undertaken in the automatic derivational morphology field \cite{Lovins68,Porter80}. Consequently, several tools can, for a given term, provide a list of candidates-derivatives likely to belong to its derivational family. Some of these systems are based upon derivation learning \cite{SnoverAl02}, whereas others apply general and ad hoc rules to generate derivatives \cite{Namer03}. This research employs a probabilistic system that searches the term's stem and attaches successively to that stem all the suffixes it knows in order to return a derivational list for this term \cite{Gaussier99}. This tool is based on stemming and suffixation learning from an inflectional lexicon. It meets the requirements of the method described above: the stemming learning parameters can be set up more or less strictly, and the weakest constraints make it possible to generate so many candidates-derivatives that the whole of the derivational family is created, or almost -- valuing recall over precision since the noise filtering happens after the derivatives are generated.

Following this method, each significant entry (nouns, verbs, adjectives, adverbs) was addressed by means of the French dictionary used for the filtering process (the \textit{Dubois} dictionary, see below) with the generation tool. For each entry a list of candidates-derivatives was obtained, covering at best the derivational field of this entry, and more besides. It should be noted that entries shorter than 3 syllables have been ignored by the method, because the tool cannot find a stem for shorter words, and it is absolutely necessary for the suffixation process. Another restriction is applied on generated forms: each proposition is compared to a lexicon extracted from a large corpus (5 years of \textit{Le Monde} newspaper, 100 millions words) in order to eliminate chim{\ae}ras and nonexistent words from the derivational resource.

Furthermore, the \textit{Dubois} dictionary utilised is a general electronic French dictionary that contains derivational information for each entry \cite{dubois-duboischarlier97}. The dictionary is made up of 2 computer files respectively dedicated to the description of verbs (12,309 verbal entries) and of other words (102,917 non-verbal entries) in French\footnote{In order to make the text clearer, we designate the whole dictionary by the name \textit{Dubois}, and the two parts respectively by \textit{Dubois of verbs} and \textit{Dubois of words}.}. As shown in the table \ref{tbl:DuboisVex}, the \textit{Dubois} dictionary contains very rich and varied information, particularly in the verb component, which is considerably more detailed (conjugation, syntactic schema\dots). A specificity of the \textit{Dubois} lies in providing all the information types for each meaning, which is more rigorous than most dictionaries tend to be. Information types concern semantics (domain, class, sense), syntax (operator, syntactic construction\dots) and morphology (conjugation, derivatives, name). Construction and conjugation fields only appear in verbs, and each type of information is consistent in the two parts of the \textit{Dubois} dictionary.

\begin{table}[ht]
\centering
\scriptsize{
\begin{tabular}{|l|p{2.8cm}|p{2.8cm}|}
\hline
{\rm \textit{Lemma}}         &formaliser 01(s)                               &formaliser 02\\
\hline\hline
{\rm \textit{Domain}}        &PSY                                            &MAT\\
{\rm \textit{Class}}         &P1c                                            &T4b\\
{\rm \textit{Operator}}      &sent offense D                                 &r/d formel\\
{\rm \textit{Sense}}         &se choquer, se vexer                           &donner formalisation à\\
{\rm \textit{Example}}       &On se f$\sim$ de sa conduite. Cette conduite a f$\sim$ P. &Le mathématicien f$\sim$ une théorie. Cette méthode ne se f$\sim$ pas.\\
{\rm \textit{Conjugation}}   &1aZ                                            &1aZ\\
{\rm \textit{Construction}}  &P10b0 T3100                                    &T1308 P3008\\
{\rm \textit{Derivatives}}   &1-\,- -\,- -\,-\,-\,- -\,-                     &-Q- -\,- RB-\,- -\,-\\
{\rm \textit{Name}}          &6L                                             &6L\\
{\rm \textit{Level}}         &2                                              &5\\
\hline
\end{tabular}
}
\caption{Entry \textit{formaliser} in the \textit{Dubois of verbs}.}
\label{tbl:DuboisVex}
\end{table}

The guidelines in the dictionary are provided as alphanumeric codes and are therefore easier for an automatic system to read than a human being. For example, the \texttt{1aZ} code from the conjugation field in the table \ref{tbl:DuboisVex} indicates that the corresponding sense (here the two senses for the entry \textit{formaliser} have the same conjugation code, as usual) belongs to the regular version (\texttt{a}) of the first (\texttt{1}) conjugation pattern \textit{aimer} (to love), and that the auxiliary (\texttt{Z}) for composed active tenses is \textit{avoir} (to have)\footnote{In French, two auxiliaries may be used in composed active tenses: \textit{avoir} (to have) and \textit{être} (to be), but only one is correct for a given verb and this information is needed for nonnative speakers and computers.}. In spite of this formalised aspect, some information fields cannot be used directly by a computer. In particular, derivational instructions are not explicit enough to give the opportunity to generate the right derivatives from the entry: instructions generally indicate which suffix to use, but not how to find the stem, nor whether the stem and the affixes undergo morphological changes because of their mutual influence. For example, the derivation fields in the table \ref{tbl:DuboisVex} provide a \texttt{Q} code for the second sense of \textit{formaliser} (to formalise), that indicates the existence of a verbal adjective with the suffix \textit{-é} in both positive (\textit{formalisé}, formalised) and negative (\textit{informalisé}, unformalised) forms. But the instruction is no more explicit regarding the negative prefix that could be founded on the privative \textit{a-} (with possible euphonious consonants, depending on the stem) or on \textit{in-} (with a possible consonant variation, depending on the stem). Because of this lack of precision, we had to use the derivation tool described above. Nonetheless, the instructions provided give enough information to take out incorrect candidates-derivatives.

\begin{table}[ht]
\centering
\footnotesize{
\begin{tabular}{|c|l|}
\hline
\textit{Candidates-derivatives} & \multicolumn{1}{c|}{Dubois' \textit{instructions}} \\
\hline\hline
coup (knock)               & -- \\
\textbf{coupure} (cut)     & nominal derivative in \textit{-ure} \\
coupable (guilty)          & -- \\
\textbf{coupage} (cutting) & nominal derivative in \textit{-age} \\
\textbf{coupant} (sharp)   & verbal adjective in \textit{-ant} \\
\textbf{coupeur} (cutter)  & nominal derivative in \textit{-eur} \\
\textbf{coupé} (cut)       & verbal adjective in \textit{-é} \\
coupon (remnant)           & -- \\
\dots                      & \dots \\
\hline
\end{tabular}
}
\caption{Filtering candidates-derivatives produced by the derivation tool.}
\label{tbl:exIndex}
\end{table}

It is quite easy to filter out wrong candidates-derivatives, by comparing the affixal characteristics of each candidate for a term with all the instructions in the derivative field of the corresponding entry: the suffix identifies the candidates that are well conformed. In the left hand column, the table \ref{tbl:exIndex} shows some of the candidates-derivatives generated by the derivation tool for the entry \textit{couper} (to cut). The bold font indicates the candidates that matched a derivational instruction (in the right hand column) in the dictionary. These candidates are thus considered as real derivatives for the current entry, and the candidates whose suffix does not match the derivational instructions are deemed erroneous, and deleted from the derivational list.

When the 115,226 \textit{Dubois}' entries were submitted to the derivation tool, about 2 million candidates-derivatives were returned. Among those candidates, 502,429 were identified as real derivatives by our methodology, i.e. about 5 derivatives per entry on average. An evaluation of these derivatives was then undertaken. Randomly taking 10,000 derivatives from the derivational resource just created, only 24 wrong creations were identified, i.e. precision was at 99.76\%. The wrong derivatives were generally created on the basis of a long original term for which the derivation tool found two different plausible stems. For each suffixation, two derivatives were generated every time, one for each stem. For example, the noun \textit{compartiment} (compartment) produced two stems, which in turn were used to generate two candidates: \textit{\textbf{compartiment}able} (compartmentalisable) and *\textit{\textbf{compar}able} with the same suffixation. Since in our method candidate control is based on the suffix, false stemming cannot be corrected, or even detected automatically. Nonetheless, the very low error rate should entail an insignificant quantity of noise in a Knowledge Management application.

However, the derivational field in the dictionary gives instructions for 542,296 derivations, which leaves out a further 39,867 derivations. The omission of these derivations is accounted for by the derivatives created by prefixation, which is not assumed by the derivation tool. Consequently, neither negative forms, nor other prefixations can be generated in the current state of this approach, unless we use another derivation tool. It can, however, be noted that when derivatives are created by prefixation, the derivation process causes a larger lexico-semantic variation in relation with the original term than does suffixation, particularly in the case of a negative prefix.

\section{From expansion to rephrasing}

This derivation-filtering method is at the origin of a very rich and precise derivational resource, which is particularly useful for query expansion. However, even if the semantic link between the members of a derivational family is effective, it is not stable between all the meanings of every member of the family. For example, in the entry \textit{formaliser} (formalise, see table \ref{tbl:DuboisVex}), the derivational field differs between the two senses involved, and among the derivational family for the entry, some derivatives are related to one sense and not to the other: the \texttt{B} code gives an instruction to generate \textit{formalisation} (formalisation), that corresponds to sense 2 (formalise) of the entry and not to sense 1 (take offence). Thus, even if derivation is a morphological process, some semantic constraints have to be taken into account when it is used in Knowledge Management. Consequently, the use of all the derivatives proposed by the derivational resource for utterance expansion is likely to throw up some inappropriate meanings, and then some noise. However, derivational instructions are displayed only for the corresponding senses in the \textit{Dubois} dictionary. In view of this peculiarity, it seemed necessary to take into account the original sense of every term so that only derivatives matching the derivational instructions for this sense were produced.

The issue is to identify the sense of the term in the utterance needing expansion, and then to select the derivatives suggested by the derivational field matching the sense. In the perspective of Information Extraction and synonymic expansion, we designed a Word Sense Disambiguation system based on syntactic analysis and applying disambiguation rules extracted from the \textit{Dubois} dictionary \cite{Jacquemin04a,Jacquemin04b} . Lexical, syntactic and semantic information provided by the \textit{Dubois} made it possible to create rules for every sense of each entry. Each type of information is converted into dependencies, terms and features schemas relative to the corresponding sense. For example, the entry \textit{prendre} (to take) has for its sense "to escape" an example field containing the sentence \textit{il prend la fuite} (he takes flight). This sentence produces a disambiguation rule that selects the sense "to escape" for the word \textit{prendre} when its direct object is the word \textit{fuite} (flight). These rules can match (or not) dependencies between words extracted by the XIP parser \cite{Roux99,AitAl02} from the utterances to disambiguate. So nearly 45\% of the significant words in submitted texts can be disambiguated, by associating the polysemantic words to one of their meanings in the dictionary. For these disambiguated terms, only the derivatives that match the derivational instructions for the selected sense may be used for expanding the text. For monosemantic terms and terms for which no disambiguation rule worked, the derivational expansion cannot be specified. Thus all the derivatives for a term in our derivational resource are used for expansion.

Moreover, when the WSD method is applied to a sentence for text expansion, the syntactic analysis performed by XIP produces a dependencies structure. This structure offers great advantages. All the syntactic dependencies constitute in one way or another a formal representation of the parsed sentence, since on the one hand the dependencies describe evenly the links between the words of the sentence, and, on the other, the lexical units in the sentence are identified as the arguments of the dependencies and their linguistic characteristics are expressed as features based on the arguments. The formal representation is propitious for standardisation of the word contents (lemmatisation, normalisation) and of the structure. Lexical and syntactic information is expressed in an optimised way to store data within a database, where it is indexed and easy to retrieve. In this form, it is easier to match information from a query with information from text containing the answer: it is associated if their respective structures coincide.

Syntactic structure also makes it possible to remedy a weakness in the derivational expansion method. In spite of the real meaning closeness generally observed between derivatives from the same derivational family, and in spite of the semantic subgroups established in the derivational family in order to ascribe the derivatives selection to the ones with the same meaning as the original term in context, the sense challenge inherent to the derivation phenomenon has not yet been overcome: members of the same derivational family show meaning variations in relation to the nature of the suffix used, but above all because of the rewording from a lexical category into another \cite{HathoutTanguy02}. The syntactic structure of the utterance itself cannot deal with a simple expansion by a disjunctive list of derivatives, even if their sense is similar. For example, the sentence \textit{il a coupé le courant} (he cut off the power) can be expanded by a derivative \textit{coupure} (power cut) coming from \textit{couper} (to cut off). But the corresponding utterance *\textit{il a coupure le courant} (he power cut the power) is unsatisfactory. In order to build a correct expanded utterance, successively replacing the original terms by a list of their derivatives is not enough: it is necessary to rephrase the sentence. This action must be taken on the syntactic structure of the original sentence: the structure must be modified in such a way that a derivative can be substituted for the original term in the sentence without rendering it ungrammatical. The syntactic dependencies structure produced during the word sense disambiguation process provides the opportunity to simulate the rephrasing through the dependencies structure in order to avoid the generation issue.

\begin{figure}[ht]
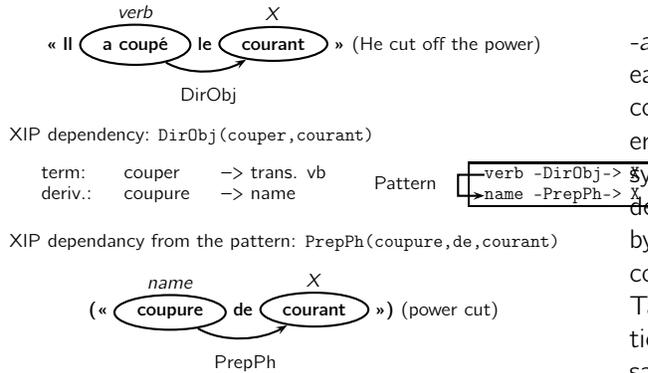

\begin{center}
\begin{minipage}{0.45\textwidth}\scriptsize{
\begin{center}\textbf{«~Il \ovalnode{couper1}{a coupé} le \ovalnode{courant1}{courant}~»} (He cut off the power)\end{center}\vspace{3mm}
\nccurve[angleA=330,angleB=210]{->}{couper1}{courant1}\nbput{DirObj}
\nput[labelsep=2pt]{90}{couper1}{\textit{verb}}\nput[labelsep=2pt]{90}{courant1}{\textit{X}}

XIP dependency: \texttt{DirObj(couper,courant)}\\[2mm]
\begin{tabular}{lll}
 \begin{tabular}{lll}
  term:       & couper  & --> trans. vb \\
  deriv.: & coupure & --> name
 \end{tabular}
 & Pattern &
 \begin{tabular}{|c|}
  \hline
  \rnode{X1}{\texttt{verb --DirObj--> X}} \\
  \rnode{X2}{\texttt{name --PrepPh--> X}} \\
  \hline
 \end{tabular}
\end{tabular}\ncbar[angle=180]{->}{X1}{X2}\\[2mm]

XIP dependancy from the pattern: \texttt{PrepPh(coupure,de,courant)}\\
\begin{center}\textbf{(«~\ovalnode{couper2}{coupure} de \ovalnode{courant2}{courant}~»)} (power cut)\end{center}}
\nccurve[angleA=330,angleB=210]{->}{couper2}{courant2}\nbput{PrepPh}
\nput[labelsep=2pt]{90}{couper2}{\textit{name}}\nput[labelsep=2pt]{90}{courant2}{\textit{X}}
\end{minipage}
\end{center}
\caption{Rephrasing into a dependency structure with the help of a derivation pattern}
\label{fig:reformulation}
\end{figure}

Ideally, an automatic system is needed that can easily and correctly rephrase a sentence such as \textit{il a coupé le courant} (he cut off the power) as \textit{la coupure de courant} (the power cut). However, text generation is still a research issue confronted with tricky problems in morphology, syntax, semantics and even pragmatics. However, if the dependencies structure coming from the morpho-syntactic analysis of the original sentence constitutes a standardized representation, the same is true of the reformulation. Therefore, it is possible to rephrase an utterance virtually without generating a real sentence: one only has to build the dependencies structure where the dependencies are the same as the ones that would have been produced by an XIP analysis of the rephrased sentence had it been generated. Thus the issue is to build a correct new dependencies structure from the original one. Designing syntactic derivation patterns that would make it possible to induce the derived dependencies structure from the original one was also considered as part of this research. Figure \ref{fig:reformulation} shows the simulated rephrasing process: the original sentence is processed by the morpho-syntactic analyzer XIP and syntactic dependencies are extracted. Word Sense Disambiguation rules are applied to select the contextual meaning of the terms (not shown on figure \ref{fig:reformulation}) in order to establish the correct derivatives. A derivation pattern depending on the original syntactic structure, on the category of the original term and on one of the derivatives is applied in order to create a new syntactic structure where the derivative is an argument instead of the original term. The new structure corresponds to the XIP analysis of the rephrased utterance (simulated in brackets) that should not be \textit{effectively} generated.

It was determined that a derivational XIP grammar should be created in order to simulate correct derivational rephrasing in most cases. For this purpose, the derivational rephrasing process was considered on a relatively large scale and in a real life environment. Certain changing parameters had to be studied: the lexical category of the original word and of the derivative, the suffix in the original word and in the derivative, and for verbs, the syntactic schema. By successively varying the value of these parameters, all the possible combinations of authentic original texts and corresponding sentences rephrased by the research team were duly tested. For each combination, 3 instances were randomly selected from among the \textit{Dubois} dictionary entries (for example, 3 direct transitive verbal entries with the \textit{-iser} suffix that comprise instructions for a nominal derivative with the \textit{-ation} suffix). By successively questioning Google with each entry as a request, the first 20 different phrastic contexts where the entry appeared were chosen. Every selected sentence was then submitted to morpho-syntactic analysis by XIP in order to extract syntactic dependencies. The original sentence was also rephrased by using the derivative corresponding to the parameters combination, and the new sentence submitted to XIP. Taking into account the recurrence of an initial syntactic schema (at least 5 occurrences for every entry in the same parameters combination) and the regularity of the corresponding dependencies structure in the rephrasings (at least 2/3 of the instances of the recurrent initial syntactic schema are rephrased into the same dependency structure), 54 derivation patterns were drawn such as the one shown in figure \ref{fig:reformulation} including 34 patterns for a derivation from a verb.

The derivation patterns were tested by rephrasing the sentences from a corpus to as great an extent as possible. The corpus was drawn from a general encyclopedic dictionary, the \textit{Encyclopédie Hachette Multimédia} \cite{AlcouffeAl00b}. This corpus contains 50,000 words from articles with the tag \textit{Roman Antiquity}. The corpus was morpho-syntactically analyzed and submitted to Word Sense Disambiguation in order to select derivatives that could be used for rephrasing. From this result, 807 derivative patterns were applied to reformulate sentences. In order to evaluate the quality of the new dependencies structures, we generated sentences where the selected derivative took the original word's place, modifying the syntactic structure to keep the sentence well-formed, and submitted the new sentences to XIP analysis. For 656 reformulations (81.29\%), the dependency produced by the derivative pattern matched the XIP analysis of the sentence as originally written. The non-matching cases were due mainly to errors in the part-of-speech tagging of the original word (102 occurrences, 12.64\%) or to syntax analysis in either the original or the rephrased sentences (37 occurrences, 4.58\%). Only 12 errors (1.49\%) may be legitimately attributed to the derivative patterns, when the original sentence has a particular syntactic schema.

\section{Rephrasing evaluation in a question answering task}

\subsection{Derivational rephrasing in a QA system}

This research has thus produced a rich and precise derivational lexicon that will associate to a word's specific sense only those derivatives with a similar meaning. A method that can rephrase utterances through morphological derivation of a term was also developed, which takes into account both the original term meaning when proposing derivatives and the syntactic structure of the rephrased utterance. This method simulates the rephrasing into a dependencies structure in order to avoid the text generation issue. The next step is thus to integrate the method in a question answering system and to supply more textual formulations in order to match elements from the question and from the answer. Since a major issue in question answering is how to match texts with an identical meaning but a different wording, a derivational rephrasing module should help the existing synonymic rephrasing module in the question answering system employed in this research.

\begin{figure}[ht]
\centering
\includegraphics[width=0.45\textwidth]{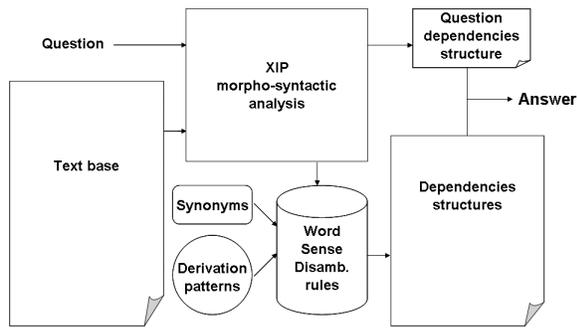}
\caption{The architecture of our question answering system}
\label{fig:qa}
\end{figure}

The question answering system developed \cite{Jacquemin05} employs an original methodology to find textual answers to a question by matching dependencies structures. Such structures are extracted by morpho-syntactic analysis of both question and text; then Word Sense Disambiguation is performed in order to select correct synonyms according to the initial meaning. It is then possible to simulate a synonymic rephrasing by enriching the dependencies structure. A feature of this approach is the special deep pre-processing undergone by the text base instead of the question. The method uses only minimal analysis to extract dependencies from the question. This approach is connected with the fact that XIP is better at analysing normal text than questions, and above all it is related to the necessity to have as much syntactic context as possible in order to improve the Word Sense Disambiguation results \cite{Weaver49,Reifler55}. The classical approach in query expansion was improved by rephrasing performed on the texts, first through synonymy and subsequently through derivation. The search for an answer is performed by comparing the question minimal structure with the text enriched structures, and matching the inner dependencies (see figure~\ref{fig:qa}).

\begin{figure}[ht]
\centering
\begin{minipage}{0.45\textwidth}\footnotesize{
\begin{center}\textbf{«~De quel chef Domitien est-il le successeur?~»}\\(Of which chief is Domitian the successor?)\end{center}
\textit{Question's structure:}\\[1mm]
\texttt{ \rnode{Q2}{PrepPh(successeur,de,chef)}\\
         \rnode{Q1}{ATTRIBUTE(Domitien,successeur)}}\\[3mm]
\textit{Text's structure:}\\[3mm]
\texttt{\begin{tabular}{|l|p{6cm}}
 \cline{1-1}
 SUBJECT(succéder {\red OR remplacer},Domitien)                & \\
 ATTRIBUTE(empereur {\red OR chef},Titus)                      & \textrm{Base dependencies} \\
 DirObj(succéder {\red OR remplacer},empereur {\red OR\ chef}) & \\
 \cline{1-1}
\end{tabular}}\\[2mm]
\texttt{\begin{tabular}{|l|p{6cm}}
 \cline{1-1}
  \begin{tabular}{l}
  \rnode{R1}{ATTRIBUTE}(Domitien,successeur) \\
  \rnode{R2}{PrepPh}(successeur,de,empereur {\red OR chef})\\
  \end{tabular}                                              & \textrm{Derivational dependencies} \\
 \cline{1-1}
\end{tabular}}\\
\nccurve[angleA=180,angleB=180]{<->}{Q1}{R1}
\nccurve[angleA=180,angleB=180]{<->}{Q2}{R2}
\begin{center}\textbf{«~\dots Domitien succéda à l'empereur Titus\dots~»}\\(Domitian succeeded to the emperor Titus)\end{center}
}\end{minipage} 
\caption{Questioning a dependencies structure with synonymic and derivational rephrasing}
\label{fig:interro}
\end{figure}

Since the derivational method employed in this research also uses XIP morpho-syntactic analysis and the Word Sense Disambiguation system to collect information from an utterance and to simulate rephrasing with the same meaning, it seemed natural to integrate it into the question answering system. Figure \ref{fig:interro} shows the mechanism of the question answering system. A minimal morpho-syntactic analysis is performed on the question in order to extract the dependencies structure (\textit{Question's structure}) that has to be matched with the text enriched structures. Furthermore, the text base to question has been pre-processed: morphological, syntactic and semantic analysis as well as rephrasing are performed before the request phase. The morpho-syntactic analysis produces the \textit{base dependencies} corresponding to the sentence structure. When the Word Sense Disambiguation rules have been applied to the terms in the syntactic structure, both synonyms and derivatives that match the original senses are selected to perform rephrasing: synonyms are inserted into the existing dependencies (in red), disjunctively to the corresponding original terms that belong to the same lexical category; and for derivatives, the corresponding derivation patterns are applied in order to create new dependencies structures simulating rephrasing (\textit{derivational dependencies}). Answering the question consists in returning sentences from the text that contain the same data as does the question. In figure \ref{fig:interro}, the question is answered by matching its structure with dependencies from a text structure. All the matching dependencies in the text come from derivational and synonymic rephrasing.

The current version of the question answering system developed in this research programme cannot be entered  in competitions like TREC (\textit{Text REtrival Conference}, \cite{Harman92,VoorheesBuckland05}) or CLEF (\textit{Cross-Language Evaluation Forum}, \cite{PetersAl02,PetersAl04}). On the one hand the system cannot select precisely the answer to the question since no such module has yet been developed to perform this selection: the answer to the question appears in a full sentence. On the other hand, the system is currently based on one reference dictionary, the \textit{Dubois}, that exists only for French: the rephrasing methods, and consequently the answering process, can only be applied to French questions and texts. Furthermore, lack of time and of human resources prevented the organisation and evaluation of this system on a larger scale. However, considerable efforts were made to measure impartially the efficiency of the question answering system and the impact of derivational rephrasing on the results. The TREC 8 campaign \cite{Voorhees99} evaluated question answering systems for English and the evaluation design has characteristics very close to an experiment that this research was indeed able to implement. In this evaluation, 200 questions are submitted to systems, which have to return up to 5 answers, sorted by relevance. All the questions have at least one correct answer in the text base, and a correct answer should appear in a 50 words window. A score is assigned to each question, depending on the inverse rank of the first correct answer: the score is 1/1 if the first answer is correct, 1/2 if the second answer is correct and the first one is wrong, 1/3 if the third answer is correct and the first two are wrong and so on until the fifth answer. The global score for a system is the mean of every question's scores.

\begin{table}[ht]
\centering
\footnotesize{
\begin{tabular}{|l|c|c|}
\hline
\textit{Rephrasing levels}                & \textit{Score}  & \textit{No answer} \\
\hline\hline
\textit{Baseline}                         & 0.295           & 139 \\
\textit{Base rephrasing}                  & 0.462           & 105 \\
\textbf{\textit{Derivational rephrasing}} & \textbf{0.467}  & \textbf{104} \\
\textit{All the enrichments}              & 0.504           & 97 \\
\hline
\end{tabular}}
\caption{Evaluation results}
\label{tbl:eval}
\end{table}

The text base questioned is drawn from 500 articles with an \textit{Antiquité} (Antiquity) tag extracted from the \textit{Encyclopédie Hachette Multimédia}. After reading all the articles and without the texts in front of them, 8 people from outside the project proposed 25 questions each (i.e. a total of 200 questions as in TREC 8) about information content in the texts. All the questions are in correct French. The answers are full sentences, which seemed more relevant than a 50 words window. In order to highlight the real influence of the derivational rephrasing in the answering process, the system was made to question the texts at several levels of pre-processing (table~\ref{tbl:eval}): for the \textit{baseline}, only the significant terms (nouns, verbs, adjectives, adverbs) were stored in an index; for the \textit{base rephrasing}, the base structure is extracted by an XIP analysis and a first synonymic rephrasing is performed with the few synonyms coming from the \textit{Dubois} dictionary; the \textit{derivational rephrasing} corresponds to the \textit{base rephrasing} with the derivational rephrasing method described above; the highest level of rephrasing includes \textit{all the enrichments}, i.e. a derivatives structure that contains the derivational rephrasing, the synonymic rephrasing with synonyms that come from several dictionaries (\textit{Dubois}, \textit{EuroWordNet}, \textit{Bailly}, \textit{Memodata}) and a pronominal co-reference procedure.

\subsection{Results and discussion}

Further to this evaluation, despite the quality of the derivational resource created and in spite of the capacity of this particular method to simulate grammatical derivational rephrasings of texts very close to their original meaning, it can be observed that this enrichment does not greatly improve the results achieved. The derivational rephrasing provides only one more answer. However, no answer would be found for this question without the derivation process. Moreover, the proposed answer is correct and first-ranked for the question (see figure~\ref{fig:interro}). It is also remarkable that the derivation process did not damage the results \cite{ClarkeAl00,Monz03}. By examining the system performance in greater detail, as much in the successful answers as in the weaknesses, certain error explanations and several ways to improve the system were identified.

Firstly, at least 11 cases were noted as being without answer where an idea was expressed with a verbal construction in the question and with a nominal or adjectival expression in the text. At this point, all the rephrasing processes are applied to texts and none to the questions. The exceptional wealth of information contained in the \textit{Dubois of verbs} can be confirmed; the \textit{Dubois of words} is not as complete, and the derivation field is often poorer than in the verbal part: in the case of a verbal entry, all meanings are drawn together, providing instructions for the whole derivational family, whereas the nouns, adjectives or adverbs sometimes have omissions, and do not provide instructions by means of which the corresponding verb, adjective, adverb or noun may be found. Consequently the derivational rephrasing is incomplete, and no match can be made with a missing derivative that appears in the question. Thus  the gaps in the \textit{Dubois words} derivation fields must be filled in by symmetrising the derivation instructions from the \textit{Dubois verbs}. Semantic fields like \textit{Domain} (see table \ref{tbl:DuboisVex}), that are consistent in the two parts of the dictionary, should share the instructions between the senses. Secondly, in 8 cases neither the derivational rephrasing, nor the synonymic rephrasing could simulate the question formulation, and thus provide an answer, because in the questions the two types are combined: the question addresses the same notion as the text, but the part of speech and the word are occasionally different. Thus derivational and synonymic rephrasing should also be combined, by derivational rephrasing after synonymic rephrasing or by synonymic rephrasing after derivational rephrasing (or both).

The implementation of these propositions might bring a significant improvement to the system. It should provide correct answers to the 19 marked questions that did not get any answer from the current system. If this proves to be the case, the results would be improved by nearly 10\%. Finally, a small test was undertaken using the derivational resource created to perform a classical derivational expansion on 5 articles from the corpus. Five questions from the evaluation were posed, whose answer was in one of these articles. These questions were correctly answered at the \textit{derivational rephrasing} level of the evaluation. In this test, a dramatic reduction in successful performance ensued in that two questions did not generate any correct answer, and only one had the correct answer in the first place. The mean score for these questions is 0.367 (to 0.767 with the derivational rephrasing). This scaled-down test was too small to be strictly interpreted, but it shows a distinct tendency of the approach taken by this research to preserve the quality of the results, contrary to the classical derivational expansion, which usually impacts negatively upon accuracy \cite{Hull96}.

\section{Conclusion}

In textual Knowledge Management disciplines, and more specifically in question answering, the different means of expressing the same information in a sentence can be a major source for identifying the meaning of the contents. Classical semantic relations such as synonymy or hyperonymy often provide new wordings in most of the current approaches, but the part-of-speech variations are still an issue that needs to be worked on, especially through morphological derivation.

The combined use of a derivation tool with few constraints for a very large recall, and a general French dictionary that provides derivational guidelines made it possible to create a derivational French lexicon that is particularly rich and precise. Moreover, the specific lexico-semantic characteristics of the \textit{Dubois} dictionary - mainly the systematic association of the derivational guidelines with the corresponding meaning - and the Word Sense Disambiguation process developed by this research combine to provide access to derivatives with a close meaning for a term in a selected sense. By using the XIP morpho-syntactic analyser to apply Word Sense Disambiguation rules, a syntactic dependencies structure that constitutes a formal representation of the utterance was extracted for each disambiguated utterance. In this structure it is possible to simulate derivational rephrasing of the original sentence: applying some derivation patterns leads to designing new dependencies involving the proposed derivative; these represent a rephrased utterance without generating it.

The derivational rephrasing process was integrated with the question answering system in order to measure its quality and impact on performance. The evaluation design followed for the French language copies the question answering track used in the TREC 8 competition. In spite of the modest results increase due to the derivational rephrasing method employed, observation confirms that it never damages performance in the way derivational expansion usually does. Moreover, following careful analysis of the results of the questions as well as those of the questioned texts, some promising ideas emerged to enable system improvements, notably by enriching the dictionary's derivational information field of non-verbal entries (symmetrisation from the verbal entries), and by performing a derivational rephrasing on the dependencies after the synonymic enrichment application, or by performing a synonymic rewording on the dependencies after the derivation patterns application, or both.  Plans are currently being advanced to investigate the opportunity to integrate this procedure into the QALC question answering system \cite{deChalendarAl02}, based on deep processing of the questions and working on French language \textit{inter alia}.

\bibliographystyle{apalike-fr}
\bibliography{bj}

\end{document}